\documentstyle[psfrag,aps,preprint,epsfig,axodraw]{revtex}

\begin{document}

\psfrag{mt}{$m_t~(\mbox{GeV})$}
\psfrag{1+ru}{$1+r_U$}
\psfrag{0.01}{$0.01$}
\psfrag{0.1}{$0.1$}
\psfrag{1}{$1$}
\psfrag{0}{$0$}
\psfrag{50}{$50$}
\psfrag{100}{$100$}
\psfrag{150}{$150$}
\psfrag{200}{$200$}

\tighten

\preprint{TU-694}

\title{Singular K\"{a}hler Potential and Heavy Top Quark  \\
in 
A Democratic Mass Matrix Model}
\author{
Mitsuru Kakizaki\footnote{e-mail: kakizaki@tuhep.phys.tohoku.ac.jp} 
and Masahiro Yamaguchi\footnote{e-mail: yama@tuhep.phys.tohoku.ac.jp}}
\address{Department of Physics, Tohoku University,
Sendai 980-8578, Japan}
\date{\today}
\maketitle
\begin{abstract}
It is pointed out that heavy top quark mass can be attributed to  
a singular 
normalization of its kinetic term, in which rescaling into canonical 
one yields large top Yukawa coupling. We pursue this novel 
possibility in a democratic mass matrix model where only the normalization 
of  the third generation
can be different from that of the other two generations.
With diagonal breaking of democratic $S_3$ symmetry, we show that the 
singular normalization for the top quark is essential to reproduce observed
quark masses and mixing angles.
We also briefly argue other applications of this mechanism.
\end{abstract} 

\clearpage

\section{Introduction}

The origin of the flavor structure is a big puzzle in particle physics. 
In the standard model, the masses and mixing of quarks and leptons are 
just parameters given by hand. Obviously this is unsatisfactory, and there
have been many attempts to explain the fermion mass matrices in the literature
\cite{Froggatt:1979nt,Dine:1993np,Pouliot:1993zm,Kaplan:1993ej,Ibanez:ig,Pomarol:1995xc,Hall:1995es,Arkani-Hamed:1999dc}.

There are some characteristic features in the quark and lepton mass
matrices.  As for the mixing among different generations, the quark
sector has all small mixing angles whereas the neutrino sector has two
large mixings and one small one. Another important property concerns
the mass hierarchy of the quarks and leptons. The third generation of
the quarks and charged leptons are heavier than the other generations.
In particular, the top quark is much heavier than the others.

Here we shall point out that
the large top quark mass can be accounted for by a singular
normalization of its kinetic term. After rescaling the fields to get
the canonical normalization, one can obtain a large Yukawa coupling
and thus a large mass of the top quark.  We shall demonstrate this
interesting possibility in the framework of a democratic fermion mass
matrix model
\cite{Harari:1978yi,Koide:1983qe,Fritzsch:1995dj,Fukugita:1998vn,Fujii:2002jw,Akhmedov:2000yt}. 
In the democratic $S_3$ symmetry, the fermions belong to 3
dimensional reducible representations, which are decomposed into 1 and
2 dimensional representations.  The trivial representation is
identified with the field in the third generation. 
This group theoretical structure allows the third generation to have different
normalization from the other two generations\cite{Hamaguchi:2002vi}. 
We will show that, with diagonal breaking of the $S_3$ symmetry, the singular normalization of the top quark is
essential to reproduce observed quark masses and mixing angles.
We will also discuss other examples where this mechanism may be important. 

\section{The Model}

The model we are considering was described in Refs. 
\cite{Harari:1978yi,Koide:1983qe,Fritzsch:1995dj,Fukugita:1998vn,Fujii:2002jw,Akhmedov:2000yt}. 
Non-canonical
K\"ahler potential was introduced in Ref. \cite{Hamaguchi:2002vi}. 
We would like 
to briefly review essential points of our model. Here we focus on the
quark sector. See Refs. \cite{Fritzsch:1995dj,Fukugita:1998vn,Fujii:2002jw,Akhmedov:2000yt} for the successful extensions to the lepton sector, including
neutrino masses and mixing.

In the quark sector, there are a product of permutation symmetry
groups $S_3 (Q_L) \times S_3 (U_R) \times S_3 (D_R)$. Under $S_3
(Q_L)$, the doublet quarks $Q_{Li}$ ($i=1,$ 2, 3) transform as 3
dimensional representation, ${\bf 3}$, which is decomposed into two
irreducible representations: ${\bf 3}={\bf 1}+{\bf 2}$. Here ${\bf 1}$
is a trivial representation and ${\bf 2}$ is two dimensional one.  In
fact, $(Q_{L1}+Q_{L2}+Q_{L3})/\sqrt{3}$ does not change under the
permutation. The other two combinations will constitute a basis of
${\bf 2}$. Similar arguments can apply for $SU(2)_L$ singlet quarks
$U_{Ri}$ and $D_{Ri}$.

In the following we consider the minimal supersymmetric standard model (MSSM)
to illustrate our points, and  use the terminology of supersymmetry,
such as K\"ahler potential and superpotential. Supersymmetry is,
however, not essential for our subsequent argument.

To begin with we will develop some formalism. 
Let us introduce a ${\bf 3}$ representation, $X_i$, 
which will be identified with three families of the MSSM matter fields later.
The $S_3$ invariant K\"{a}hler potential is written
\begin{eqnarray}
  K = [Z^X_I I + Z^X_J J]_{ij} X^\dag_i X_j, 
  \label{eq:kahler}
\end{eqnarray}
where
\begin{eqnarray}
  I \equiv \left( 
    \begin{array}{ccc}
      1 & 0 & 0 \\
      0 & 1 & 0 \\
      0 & 0 & 1
    \end{array}
  \right), \quad 
  J \equiv \frac{1}{3} 
  \left( 
    \begin{array}{ccc}
      1 & 1 & 1 \\
      1 & 1 & 1 \\
      1 & 1 & 1
    \end{array}
  \right).
\end{eqnarray}
Reflecting the fact that the above K\"{a}hler potential is
a bilinear function of $X_i$,
there arise two invariants: 
the universal matrix $I$ and the democratic matrix $J$.
Here $Z_I^X$ and $Z_J^X$ are functions of fields in general and 
we omit terms which have no relation to mass matrices.
We assume that some dynamics fixes  vacuum expectation values (vevs),
\begin{eqnarray}
  \langle Z^X_{I,J} \rangle = z^X_{I,J}.
\end{eqnarray}
Existence of the democratic part $z^X_J$ plays an essential role
in our arguments. 
In fact, the non-universal kinetic terms stem from $z_J^X$,
\begin{eqnarray}
  K^{\rm dem}_{ij} = [z^X_I I + z^X_J J]_{ij}.
  \label{eq:metric}
\end{eqnarray}
Using the following matrix:
\begin{eqnarray}
  A=\left(
    \begin{array}{ccc}
      1/\sqrt{2} & 1/\sqrt{6} & 1/\sqrt{3} \\
      -1/\sqrt{2}  & 1/\sqrt{6} & 1/\sqrt{3} \\
      0  &  -2/\sqrt{6}      & 1/\sqrt{3} 
    \end{array}
  \right),
\end{eqnarray}
$K^{\rm dem}$ is 
diagonalized as 
\begin{eqnarray}
   K^{\rm diag} = A^T K^{\rm dem} A = z^X_I I + z^X_J T,
\end{eqnarray}
where
\begin{eqnarray}
   T \equiv    \left(
     \begin{array}{ccc}
      0 & 0 & 0 \\
      0 & 0 & 0 \\
      0 & 0 & 1
     \end{array}
    \right).
\end{eqnarray}
Hereafter we will call the field basis obtained this way the diagonal basis.
The kinetic terms are written explicitly
\begin{eqnarray}
  K_{\rm kin} = \left( 
      \begin{array}{ccc}
        z^X_I & 0 & 0 \\
        0 & z^X_I & 0 \\
        0 & 0 & z^X_I + z^X_J
      \end{array}
      \right)_{ij} X^{{\rm diag}\dag}_i X^{\rm diag}_j, \quad 
  X^{\rm diag} = A^T X.
\end{eqnarray}
In this basis, it becomes clear that the reducible ${\bf 3}$
representation consists of the trivial ${\bf 1}$ representation and the
irreducible ${\bf 2}$ representation.  After rescaling the fields
by using the diagonal matrix
\begin{eqnarray}
  C_X =
  {\rm diag}\left( \frac{1}{\sqrt{z^X_I}}, \frac{1}{\sqrt{z^X_I}}, \frac{1}{\sqrt{z^X_I(1 + r_X)}} \right), \quad r_X \equiv z^X_J/z^X_I, 
\label{eq:rescale}
\end{eqnarray}
we obtain canonically normalized kinetic terms:
\begin{eqnarray}
  K_{\rm kin} = X_i^{{\rm can}\dag} X_i^{\rm can}, \quad 
  X^{\rm can} = C_X^{-1} X^{\rm diag}.
\end{eqnarray}
We will call this field basis the canonical basis.

Let us consider Yukawa interaction which consists of two matter fields
$X_L,X_R^c$ and a Higgs field $H$,
\begin{eqnarray}
  W = Y^{\rm dem}_{ij} X_{Li} X_{Rj}^c H,
\end{eqnarray}
where $Y^{\rm dem}$ represent a Yukawa coupling matrix. 
The $S_3(X_L) \times S_3(X_R^c)$ symmetry allows $Y^{\rm dem}$
to have only the democratic matrix:
\begin{eqnarray}
  Y^{\rm dem} = y_0 J,
\end{eqnarray}
where $y_0$ is a constant.
It is obvious that 
\begin{eqnarray}
  Y^{\rm diag} = y_0 T
  \label{eq:diag}
\end{eqnarray}
in the diagonal basis and that 
\begin{eqnarray}
  Y^{\rm can} = C_L Y^{\rm diag} C_R = {\rm diag} 
  \left( 0, 0, \frac{y_0}{\sqrt{z^L_I z^R_I (1 + r_L) (1 + r_R)}} \right)
\end{eqnarray}
in the canonical basis, and thus no mixing angle arises.
Notice that only the $(3,3)$ element 
in $Y^{\rm diag}$ and $Y^{\rm can}$ survives
since both of the ${\bf 1}$ 
representation of $X_L$ and the one of $X_R^c$ must be
involved in order for couplings to be invariant.
Therefore, there appear two massless fields and one massive field
after the Higgs develops a vev,
which, roughly speaking, simulates our world
\footnote{Democratic structure of Yukawa interactions can be also realized in 
brane-world models\cite{Watari:2002fd}.}.

$S_3$ breaking parameters must be involved
in order to make our model realistic.
Let us assume that the small $S_3$ breaking matrix possesses a diagonal form:
\begin{eqnarray}
  Y^{\rm dem} = y_{0} \left[
    J
     +\left(
       \begin{array}{ccc}
        -\epsilon & 0 & 0 \\
            0   & \epsilon & 0 \\
            0   & 0  & \delta
       \end{array}
      \right)
      \right].
      \label{eq:yukawa_dem}
\end{eqnarray}
Diagonalization of the kinetic terms using $A$ 
followed by rescaling the fields using $C_X$
drives the Yukawa matrix to the following form:
\begin{eqnarray}
  Y^{\rm can} =  y_{0}  C_L \left[ T+ A^T
      \left(
      \begin{array}{ccc}
        -\epsilon & 0 & 0 \\
        0 & \epsilon & 0 \\
        0 & 0 & \delta
      \end{array}
      \right) A \right] C_R
    \label{eq:yukawa_c}
\end{eqnarray}
in the canonical basis.
One is convinced that $Y^{\rm can}$ is almost diagonal 
and that resulting mixing  angles will be small.
After diagonalization of $Y^{\rm can}$,
\begin{eqnarray}
  U_L^T Y^{\rm can} U_R & = & \mbox{diag}(y_1, y_2, y_3),
\end{eqnarray}
we obtain
\begin{eqnarray}
    y_1 = Y_0 \left( \frac{\Delta}{3} - \frac{\Xi}{6} \right), \quad
    y_2 = Y_0 \left( \frac{\Delta}{3} + \frac{\Xi}{6} \right), \quad
    y_3 = Y_0 \left( 1 + \frac{\delta}{3} \right)
    \label{eq:masses}
\end{eqnarray}
where
\begin{eqnarray}
  Y_{0} & = & \frac{y_{0}}{\sqrt{z_I^L z_I^R (1 + r_L)(1 + r_R)}},\quad 
  \Xi = 2 ( \Delta^2 + 3 E^2)^{1/2}, \nonumber \\
  \Delta & = & \sqrt{(1 + r_L)(1 + r_R)} \delta,\quad 
  E = \sqrt{(1 + r_L)(1 + r_R)} \epsilon.
\end{eqnarray}
Here we choose $\delta$ and $\epsilon$ to be real 
and ignore higher orders of $\delta$ and $\epsilon$.
The unitary matrices $U_L$ and $U_R$ are given by
\begin{eqnarray}
  U_{L,R} =  \left(
    \begin{array}{ccc}
      \cos \theta & - \sin \theta & - \Lambda^{L,R} \sin 2 \theta \\
      \sin \theta & \cos \theta & - \Lambda^{L,R} \cos 2 \theta \\
      \Lambda^{L,R} \sin 3 \theta & \Lambda^{L,R} \cos 3 \theta & 1
    \end{array}
  \right),
\label{eq:ULR}
\end{eqnarray}
where
\begin{eqnarray}
  \tan 2 \theta  = \frac{\sqrt{3} \epsilon}{\delta}, \quad 
  \Lambda^{L,R} = \sqrt{1 + r_{L,R}} \frac{\xi}{3 \sqrt{2}}, \quad
  \xi = 2 \delta \left( 1 + \frac{3 \epsilon^2}{\delta^2} \right)^{1/2}.
  \label{eq:lambda}
\end{eqnarray}
Notice that in the case where $E \ll \Delta \ll O(1)$,
a hierarchical spectrum is derived:
\begin{eqnarray}
    y_1 \sim  - \frac{E^2}{2\Delta} Y_0, \quad
    y_2 \sim  \frac{2 \Delta}{3} Y_0, \quad
    y_3 \sim  Y_0.
\end{eqnarray}
Hereafter we assume this type of hierarchy since
it matches with the observed quark mass hierarchy
as we will see shortly.

Let us apply the above discussion to the quark sector in the MSSM,
in which the left-handed quark doublets $Q_i$, 
the charge-conjugated right-handed up-type quarks $U_i^c$
and the charge-conjugated right-handed down-type quarks $D_i^c$ exist.
The up-type Yukawa coupling matrix and the down-type one are diagonalized as
\begin{eqnarray}
  U_Q^{uT} Y^{\rm can}_u U_U = \mbox{diag}(y_u, y_c, y_t), \quad 
  U_Q^{dT} Y^{\rm can}_d U_D = \mbox{diag}(y_d, y_s, y_b)
\end{eqnarray}
respectively.
The Cabibbo-Kobayashi-Maskawa (CKM) matrix is given by
\begin{eqnarray}
  & & V_{\rm KM} = U_Q^{u\dag} U_Q^d \nonumber \\
  & & \qquad = \left( 
    \begin{array}{ccc}
      \cos \theta_c & \sin \theta_c & 
      - \Lambda_d^Q \sin \theta_b + \Lambda_u^Q \sin 3\theta_u \\ 
      - \sin \theta_c & \cos \theta_c & 
      - \Lambda_d^Q \cos \theta_b + \Lambda_u^Q \cos 3\theta_u \\ 
      - \Lambda_u^Q \sin \theta_t + \Lambda_d^Q \sin 3\theta_u & 
      - \Lambda_u^Q \cos \theta_t + \Lambda_d^Q \sin 3\theta_u & 
      1
    \end{array}
  \right),
\end{eqnarray}
where
\begin{eqnarray}
    \quad \theta_c = \theta_u - \theta_d, \quad
    \theta_t = 2 \theta_u + \theta_d, \quad
    \theta_b = \theta_u + 2 \theta_d.    
\end{eqnarray}
One finds that the small $S_3$ breaking parameters induce small mixing angles.
Notice that our model naturally 
includes the model discussed in Ref. \cite{Koide:1983qe}, where non-universal
kinetic terms are absent.

We can express the CKM matrix elements in terms of the mass eigenvalues.
The Cabibbo angle is determined solely by the observed values as
\begin{eqnarray}
  |V_{us}| = |\sin \theta_c| \sim \sqrt{\frac{m_d}{m_s}}
  \pm \sqrt{\frac{m_u}{m_c}}.
\end{eqnarray}
This Fritzsch relation have been known to be successful \cite{Fritzsch:1977za}.
$V_{cb}$ and $V_{ub}$ are given by
\begin{eqnarray}
  |V_{cb}|^2 + |V_{ub}|^2 & = & 
  \frac{\bar{\Lambda}_u^2}{1 + r_U}
  \left( 1 + \frac{1}{x^2} - \frac{2 \cos 2 \theta_c}{x} \right),
  \nonumber \\
  \frac{V_{ub}}{V_{cb}} & = &
  \frac{\sin (3 \theta_u - 2 \theta_c)-x \sin 3 \theta_u}
  {\cos (3 \theta_u - 2 \theta_c)-x \cos 3 \theta_u}
  \label{eq:ckm}
\end{eqnarray}
where
\begin{eqnarray}
  x & = & 
  \frac{\sqrt{1+r_D}}{\sqrt{1+r_U}} \frac{\bar{\Lambda}_u}{\bar{\Lambda}_d},
  \nonumber \\
  \bar{\Lambda}_u & = & \sqrt{(1 + r_Q)(1 + r_U)} \frac{\xi_u}{3\sqrt{2}}
  = \sqrt{(1 + r_U)} \Lambda^Q_u, \nonumber \\
  \bar{\Lambda}_d & = & \sqrt{(1 + r_Q)(1 + r_D)} \frac{\xi_d}{3\sqrt{2}}
  = \sqrt{(1 + r_D)} \Lambda^Q_d.
\end{eqnarray}
Notice that in eq. (\ref{eq:ckm}) 
values of $\theta_u$ and $\bar{\Lambda}_u$ are almost fixed
by the empirical masses as
\begin{eqnarray}
  |\theta_u| \sim \sqrt{\frac{m_u}{m_c}}, \quad 
  \bar{\Lambda}_u^2 \sim \frac{m_c^2}{2 m_t^2}
  \label{eq:mass_mixing_relation}
\end{eqnarray}
(See eqs. (\ref{eq:masses}) and (\ref{eq:ULR})).
Thus, parameters determined from $V_{cb}$ and $V_{us}$ are $x$ and $1+r_U$.
$1+r_Q$ does not play an essential role.

\section{Analysis}
Let us perform crude estimation
before making numerical analysis,
elucidating the importance of unusual structure of kinetic terms.
For the present, we adopt the following representative values:
\begin{eqnarray}
  |\sin \theta_c| = 0.22, \quad |\theta_u| = 5 \times 10^{-2}.
  \label{eq:values}
\end{eqnarray}
Taking into account that
\begin{eqnarray}
  |x| \sim O(0.1) \frac{\sqrt{1+r_D}}{\sqrt{1+r_U}},
\end{eqnarray}
the area where $|x|$ itself is close to $O(0.1)$ is preferable.
From the constraint $0.057 \leq |V_{ub}/V_{cb}| \leq 0.126$
one finds that the most favorable range is $\theta_u \theta_c > 0$ and
\begin{eqnarray}
  - 1.14 \leq x \leq - 0.61.
  \label{eq:x}
\end{eqnarray}
We discard the case where $\theta_u \theta_c < 0$ 
since rather large $|x|$ is demanded.
Combining this inequality (eq. (\ref{eq:x})) with the constraint 
$ 1.44 \times 10^{-3} \leq |V_{ub}|^2 + |V_{cb}|^2 \leq 1.94 \times 10^{-3}$,
we obtain
\begin{eqnarray}
  0.013 \leq 1 + r_U \leq 0.034.
\end{eqnarray}
for $m_c = 677~\mbox{MeV}$ and $m_t = 175~\mbox{GeV}$.
Thus, 
we conclude that considerably 
suppressed $1+r_U$ is necessary in order that our $S_3$
model with the diagonal breaking matrices 
explains the empirical masses and mixings.
Recalling that 
\begin{eqnarray}
  y_t \sim \frac{y_{0u}}{\sqrt{z_I^Q z_I^U (1 + r_Q)(1 + r_U)}},
\end{eqnarray}
this conclusion highlights a possibility that 
extremely large top mass is attributed to singular structure of 
normalization of the fields in the case of small $y_{0u}$.

\begin{figure}[ht]
  \begin{center}
    \makebox[0cm]{
      \scalebox{1.0}{
        \includegraphics{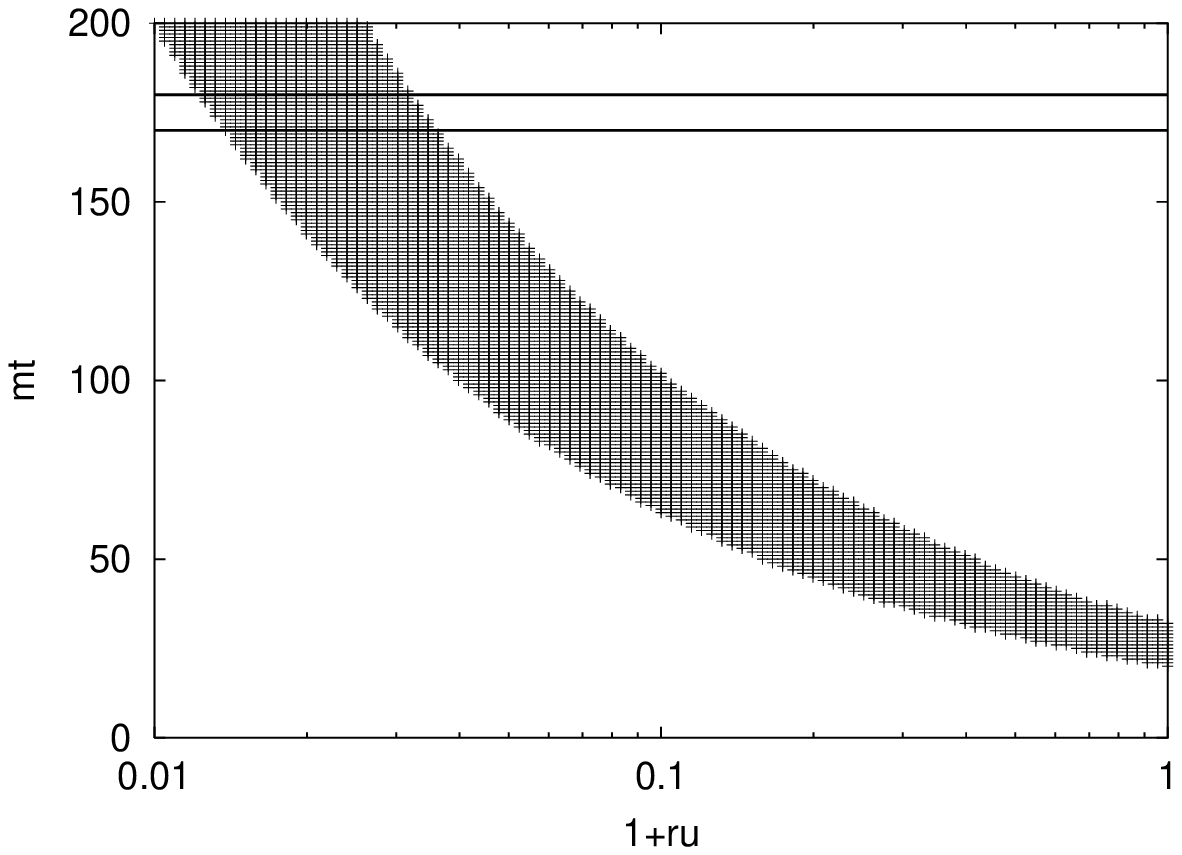}
        }
      }
    \caption{The top quark mass $m_t$ evaluated at the $Z$-boson mass scale
      as a function of $1+r_U$.
      The hatched region is allowed by the constraints 
      from $|V_{cb}|$ and $|V_{ub}|$.
      The $1 \sigma$ region of the experimentally measured top mass 
      $(170 ~\mbox{GeV}\leq m_t \leq 180 ~\mbox{GeV})$ 
      is also shown by the two horizontal lines.
      Here we use $- 1.14 \leq x \leq - 0.61$ and $m_c = 677~\mbox{MeV}$.}
    \label{fig:mt}
  \end{center}
\end{figure}

Put another way, we draw a plot of the allowed value of the top quark
mass $m_t$ as a function of $1+r_U$ when the other quantities are fit
with their experimental values as above. 
The allowed region is depicted in Fig. \ref{fig:mt}.
One finds that the allowed top mass is rather sensitive to the choice of
$1+r_U$. It is interesting to observe that 
the case of universal kinetic terms examined in Ref. \cite{Koide:1983qe} 
is completely ruled out and thus the inclusion of the singular normalization
with extremely small $1+r_U$ is essential to make the model realistic.

We now exhibit numerical analysis based on the diagonal breaking matrices.
For simplicity we set $1+r_Q = 1+r_D = 1$, which does not alter our conclusion.
The following parameter set,
\begin{eqnarray}
  & & 1+r_U = 2.53 \times 10^{-2}, \nonumber \\
  & & y_{0u} \langle H_u \rangle / \sqrt{z_I^Q z_I^U} = 27.5~\mbox{GeV}, \quad 
  \delta_u = 3.76 \times 10^{-2}, \quad
  \epsilon_u = 2.36 \times 10^{-3}, \nonumber \\
  & & y_{0d} \langle H_d \rangle / \sqrt{z_I^Q z_I^D} = 3.09~\mbox{GeV}, \quad 
  \delta_d = -4.37 \times 10^{-2}, \quad 
  \epsilon_d = 9.10 \times 10^{-3}
\end{eqnarray}
reproduces mass eigenvalues 
\begin{eqnarray}
  & & m_u = 2.26~\mbox{MeV}, \quad m_c = 683~\mbox{MeV}, \quad
  m_t = 175~\mbox{GeV}, \nonumber \\
  & & m_d = 2.47~\mbox{MeV}, \quad m_s = 94.0~\mbox{MeV}, \quad 
  m_b = 3.05~\mbox{GeV}
\end{eqnarray}
and magnitudes of the CKM matrix elements
\begin{eqnarray}
  |V_{us}| = 0.22, \quad |V_{cb}| = 0.038, \quad |V_{ub}| = 0.0036, 
\end{eqnarray}
which should be compared  with the experimental data evaluated 
at the $Z$-boson mass
scale \cite{Fusaoka:1998vc}.

we have made the comparison at the $Z$-boson mass scale. 
The same conclusion that extremely small $1+r_U$ is 
required holds even if we made this comparison at the GUT scale in the MSSM,
though the value $1+r_U$ itself is different.
This is because the effect of running comes from the top Yukawa coupling,
which appears in the wave-function renormalization of the top quark in 
supersymmetry.

Although we have so far supposed that the parameters which are responsible 
for quark masses and mixing angles are real,
 the observed $CP$ asymmetry can be also explained in this framework
by taking complex $\epsilon$ and $\delta$.

\section{Summary and Discussion}

In this paper, we have pointed out the possibility that the
wave function normalization of the top quark can be singular, which
makes it very heavy. We have illustrated this in the framework of the
democratic fermion ansatz.  There the fermion mass structure is controlled
by the $S_3$ permutation symmetries, with diagonal breaking introduced.
In this case, we can make even a stronger statement: the singular normalization
is essential to reproduce the top quark mass and the other quark masses
as well as their mixing.

This investigation presented here is indeed based on a particular form
of the $S_3$ breaking.  However, we emphasize that the large top quark
mass can be realized by invoking singular structure of the
corresponding kinetic terms.  This argument is irrespective of the
form of the $S_3$ breaking matrices.  This new idea of highly
different normalization of fields can save theories in which Yukawa
couplings would otherwise be suppressed and thus would fail to reproduce
the large top mass. Examples include
\begin{itemize}
\item Brane-world scenarios in which quark and/or Higgs fields propagate 
in the bulk.
In this case Yukawa coupling constants in four dimensions are volume-suppressed
by some powers of $M_c/M_*$, 
where $M_c$ denotes the compactification scale and $M_*$ the fundamental one;
\item $SU(6)$ GUTs. An attractive feature of the $SU(6)$ GUTs is that  the MSSM 
Higgs doublets can arise as pseudo-Goldstone 
multiplets \cite{Barbieri:1994kw}. A natural extension to $SU(6)$ from $SU(5)$
suggests to introduce ${\bf 15}$, ${\bf 6}^*$ and ${\bf 6}^{\prime *}$ 
as quarks and leptons, and
${\bf 6}$ and ${\bf 6}^*$ as Higgs multiplets. However with these matter
contents one cannot write down $SU(6)$ invariant Yukawa couplings for 
up-quark masses at the renormalizable level. Up-type Yukawa couplings arise from non-renormalizable 
operators and are 
suppressed by powers of $M_{\rm GUT}/M_*$, where 
$M_{\rm GUT}$ represents the GUT scale. Thus one conventionally introduce {\bf 20} to obtain the large top Yukawa coupling. Our mechanism of the
Yukawa enhancement due to the singular wave-function normalization, however, 
may allow us to construct a simpler model without introducing the
{\bf 20} multiplet.  
\end{itemize}

Throughout this paper, we have assumed that the singular normalization for
the top quark is realized by some means, but we have not specified a
possible mechanism. A particularly interesting possibility is that the
singular normalization is realized dynamically associated with, for
example, the electroweak symmetry breaking. Although a naive
inspection suggests that one needs a very flat potential for a field
responsible for the wave-function normalization, which is thus unlikely,
further study along this line is interesting and should be
encouraged.

\section*{Acknowledgment}               
This work was supported in part by the Grant-in-aid from the Ministry
of Education, Culture, Sports, Science and Technology, Japan, No.12047201.
MK thanks the Japan Society for the Promotion of Science for financial 
support.

\end{document}